# Excited states in hydrogenated single-layer MoS$_2$

Naseem Ud Din, Volodymyr Turkowski, and Talat S. Rahman

Department of Physics, University of Central Florida, Orlando, FL 32816-2385, USA

We calculate the excitation spectrum of single-layer MoS$_2$ at several hydrogen coverages by using a method based on first-principles Density-Matrix Time-Dependent Density-Functional Theory (TDDFT). Our results show that the fully hydrogenated system is metallic, while in the low-coverage limit the spectrum of single-layer MoS$_2$ includes spin-polarized partially filled localized mid-gap states. These states arise from s-orbitals of H atoms which make a tilted bond with the surface S atoms. The calculated absorption spectrum of the system reveals standard excitonic peaks, which correspond to the bound valence-band hole and conduction-band electron, as well as excitonic peaks that involve the mid-gap charges. As in the case of pristine single-layer MoS$_2$, binding energies of the excitons of the hydrogenated system are found to be relatively large (few tens of meV), making their experimental detection facile and suggesting hydrogenation as a knob for tuning the optical properties of single-layer MoS$_2$. Importantly, we find hydrogenation to suppress visible light photoluminescence, in agreement with experimental observations. As an aside, we contrast the effects of hydrogen coverage to that of the next two elements in the same column of the periodic table (the lightest metals), Li and Na, on the spectral properties of single-layer MoS$_2$ which lead instead to the formation of n-doped non-magnetic semiconductors that do not allow excitonic states.

## 1. Introduction

Offering versatile electronic and optical properties, two-dimensional (2D) transition metal dichalcogenides (TMDCs) have attracted much interest. They exhibit attractive properties such as strong photoresponse and transition from indirect to direct band gap as the number of atomic layers decreases to one [1-3]. Furthermore, the bandgap of 2D TMDC materials may be tuned quantitatively by varying the layer thickness and strain [4, 5]. These materials may also show edge-structure dependent semiconducting-to-metallic transitions [6]. Hydrogenation can be used as a knob to tune the structural, electronic, optical, magnetic and catalytic properties of 2D TMDCs [7]. Moreover, the adsorption of an alkali metal (K) also has significant influence on the catalytic properties of MoS$_2$ [8-11], while the exposure of Na to the MoS$_2$(0001) surface leads to expected n-type dopant shifts of the bands [12, 13]. The last result was also confirmed by a combined experimental-theoretical study [14], in which it was shown that metal adsorbates - Na and Co atoms –modify electronic structure of the surface systems, MoS$_2$(0 0 0 1) and WSe$_2$(0 0 0 1), differently. While Na atoms act as electron donors in MoS$_2$, Co atoms serve as electron acceptors (p-type) in WSe$_2$. These dopants dramatically modify the electronic structure of these materials, and thus their excitation spectrum (for example, by enhancing plasmonic properties through increased density of carriers). Another combined experimental-theoretical study[15] showed the possibility of tuning structural defects of monolayer MoS$_2$ through hydrogenation such that the system's electronic properties could shift at room temperature from the intrinsic electron (n) to hole (p) doping. In addition, hydrogenation can completely saturate the sulfur vacancies in MoS$_2$ flakes [15]. In a related TMDC, monolayer MoSe$_2$, it was demonstrated [16] that hydrogenating the system via a plasma treatment induces a charge transfer from hydrogen to MoSe$_2$ changing the electronic and excitation properties of the system (e.g., through a change of available states at different atoms), resulting in a shift of the PL peak and a ~3-time decrease of the electron mobility. The above properties can be tuned by changing temperature. For example, 500C heat treatment restores the electronic and optical properties of the system by stimulating desorption of the hydrogen atoms [16]. Recently, a quench of the photoluminescence (PL) induced by hydrogenation of single-layer MoS$_2$ was also observed [17].

Significant advances in understanding properties of hydrogenated monolayer MoS$_2$ have also been made through application of density functional theory (DFT). For example, it has been shown that electronic structure and magnetism of hydrogenated monolayer MoS$_2$ can be modified under uniaxial tensile strain [18]. In a related study, Jeon et al. showed [19] that magnetic properties of monolayer MoS$_2$ strongly depend on the hydrogen concentration and on the sites where they are adsorbed: adsorption at the center of the hexagonal ring and a relatively high concentration of H atoms produce an itinerant ferromagnetism, while adsorption on sulfur atoms and a low concentration of adatoms yield a flat-band ferromagnetism. Importantly, it was found that the electrons in the last case are distributed on H and neighbor Mo and S atoms, while in the non-magnetic case they are mostly localized on the Mo atoms neighboring hydrogen. Another important related finding is that hydrogenation significantly reduces the diffusion barriers for Ni and S complexes and helps build stable conducting nanorods on MoS$_2$ [20]. Previous DFT calculations also showed [21]that the interaction of H with MoS$_2$ becomes more favorable with increasing hydrogen concentration. Absorption of a single H per 4×4×1 cell produces a mid gap state approaching the Fermi level, and hence significantly increases the n-type carrier concentration and as result the system's electrical conductivity.

Furthermore, the interplay of spin-valley coupling, orbital physics, and magnetic anisotropy in several TMDCs (TM = Mo, W; C= S, Se) with an absorbed single magnetic transition-metal (TM) atom was explored by Shao *et al*. [22], who demonstrated that the spin-flip scattering rates in the systems strongly depend on involved orbitals, since orbital selection rules define the kinetic exchange coupling between the adatom and the charge carriers, thus opening a possibility to potential spintronic applications by tuning magnetic and spin transport properties via doping with TM adatoms. Quite remarkably, it was shown theoretically that another hydrogenated 2D material – single layer *h*-BN – demonstrates an oscillatory dependence of the bandgap on hydrogen concentration, making it possible to tune absorptive and emissive properties of the system[23].
In another DFT study of adsorptive properties of TMDCs [24] it was shown that the surface and interlaminar hydrogenations have various effects on the electronic properties of monolayers TMDCs (TM = Mo, W; C = S, Se, Te): e.g., in several systems the surface hydrogenation induces magnetism and reduces value of the bandgap, but does not modify semiconducting character of the band structure of the monolayer, while interlaminar hydrogenation induces a semiconductor-to-metal transition. These results demonstrate a potential of hydrogen functionalization of TMDCs for use in electronic and magnetic devices. Moreover, it has been shown theoretically with the example of graphene that hydrogenation might significantly enhance the critical temperature and superconducting properties of 2D materials [25, 26].

In the experimental and theoretical results described above the excitation spectrum of hydrogenated $MoS_2$ plays a very important role. It is thus somewhat surprising that there are so far no thorough *ab initio* studies of the effect hydrogenation on the excitation spectrum of $MoS_2$. To provide this much needed microscopic understanding, we have performed a combined DFT and TDDFT study of the excitation spectrum of hydrogenated single-layer $MoS_2$ by paying special attention to changes of the excitonic properties of the pure system when varying number of hydrogen atoms are absorbed in the model system. In addition, we analyze the electronic properties of single-layer $MoS_2$ under adsorption of two other atoms with a single s electron in the outer shell - Li and Na. As we shall see, only H as an adsorbate creates non-hybridized and well-isolated hydrogenic states within the bandgap, bringing forth new excitonic states that enrich the excitonic properties of pristine single-layer MoS$_2$.

## 2. Computational Details

We performed calculations based on DFT with the plane-wave and pseudopotential methods as implemented in the Quantum Espresso package. [27] We treated exchange correlation effects within the generalized gradient approximation in the form developed by Perdew–Burke–Ernzerhof (GGA-PBE) [28] and the local density approximation (LDA), as parameterized by Perdew and Zunger (LDA-PZ). [29] We used ultrasoft pseudopotentials to describe the core-valence electron interactions. For spin-orbit coupling (SOC), we treated the core electrons fully relativistically. We applied kinetic energy cutoffs of 60 Ry and 360 Ry, respectively, for calculations of the valence electron wave functions and the electron density. We used a 16×16×1 Monckhorst-Pack grid for k-point sampling of the Brillioun zone to generate a fine reciprocal-space grid. We optimized atomic positions and lattice parameters, until the residual forces converged to less than 0.01 eV/Å.

To calculate the excitonic binding energies we used the Density-Matrix TDDFT approach [30, 31], in which the Kohn-Sham equation $i\frac{\partial \Psi_{\vec{k}}(\vec{r},t)}{\partial t} = H(\vec{r},t)\Psi_{\vec{k}}(\vec{r},t)$ is solved by using the following ansatz for the wavefunction: $\Psi_{\vec{k}}(\vec{r},t) = \sum_l c_{\vec{k}}^l(t)\varphi_{\vec{k}}^{0l}(\vec{r})$, where $\varphi_{\vec{k}}^{0l}(\vec{r})$ are the static DFT wavefunctions (l is the band index, k is the wave-vector) and $c_{\vec{k}}^l(t)$ are their time-dependent coefficients. The sum in the ansatz is over all bands involved into the optical transitions. In this work, we use the two-band approximation which reduces the problem to finding bilinear combination of the c-coefficients that constitute the density matrix: $\rho_{\vec{k}}^{lm}(t) = c_{\vec{k}}^l(t)c_{\vec{k}}^{m*}(t)$. Indeed, the elements of the density matrix define practically all properties of the system - the level occupancies (diagonal elements), electronic transitions (polarization), excitons (non-diagonal elements), etc. The density matrix elements satisfy the Liouville equation: $i\frac{\partial \rho_{\vec{k}}^{lm}(t)}{\partial t} = [H(t), \rho(t)]_{\vec{k}}^{lm}$, where $H_{\vec{k}}^{lm}(t) = \int \varphi_{\vec{k}}^{0l*}(\vec{r})H(\vec{r},t)\varphi_{\vec{k}}^{0m}(\vec{r})dr$ are the matrix elements of the Hamiltonian with respect to the static wave functions (Kohn-Sham orbitals).

From the valence (v) and conduction (c) bands, one can derive the TDDFT equation for $\rho_{\vec{k}}^{cv}(\omega)$ that corresponds to the **exciton** transition by using the linear form of the Liouville equation[30]:

$$\sum_{\vec{k}'}[(\varepsilon_{\vec{k}}^c - \varepsilon_{\vec{k}}^v)\delta_{\vec{k}\vec{k}'} + F_{\vec{k}\vec{k}'}]\rho_{n,\vec{k}'}^{cv}(\omega) = E_n \rho_{n,\vec{k}'}^{cv}, \qquad (1)$$

where $\varepsilon_{\vec{k}}^c$ and $\varepsilon_{\vec{k}}^v$ are the free electron and hole energies at the specific wave vector (providing their energy dispersion), $\rho_{n,\vec{k}'}^{cv}(\omega)$ is the $N_k$-component of the polarization vector ($N_k$ is the number of points in momentum space) and n is the bound state number. The last term in the brackets in Eq.(1) is the TDDFT effective electron-hole interaction:

$$F_{\vec{k}\vec{k}'} = \int d\vec{r}_1 d\vec{r}_2 \varphi_{\vec{k}}^{0c*}(\vec{r}_1)\varphi_{\vec{k}}^{0v}(\vec{r}_1)f_{XC}(\vec{r}_1,\vec{r}_2)\varphi_{\vec{k}'}^{0v*}(\vec{r}_2)\varphi_{\vec{k}'}^{0c}(\vec{r}_2) \qquad (2)$$

defined by the exchange-correlation (XC) kernel $f_{XC}(\vec{r}_1,\vec{r}_2)$. To obtain the excitonic binding energies, we solve Eq. (1) using long-range (LR) and Slater XC kernels[31] as implemented in the BEE code that we have developed [30, 32]. We used periodic boundary conditions along x and y-axis and added a 15 Å vacuum along z-axis to eliminate the interaction between the model 2D system and its periodic images. To simulate the system with different concentrations of H, Li and Na coverage we used 1×1×1, 3×3×1, and 5×5×1 size supercells, which gave us three adsorbate concentration - full coverage, 1/9, and 1/25 coverage, respectively.

The LDA-PZ eigenenergies and $\varepsilon_{\vec{k}}^a$ and eigenfunctions $\varphi_{\vec{k}}^{0a}(\vec{r})$ were used to construct and solve the exciton eigenenergy equation (Eq. 1). The solution of Eq.(1) was obtained for the k-point set for the irreducible Brillouin zone.

Once the electronic spectrum and the exciton eigenenergies $E_n$ and the corresponding ($N_k$-component) eigenvectors $\vec{G}_n$ (with the components $G_{\vec{k},n}$ corresponding to different momentum points) were found by solving TDDFT Eq. (1), we calculated the absorption spectrum as defined below [33, 34]:

$$A(\omega) = -\text{Im}\frac{1}{\pi}\sum_n \frac{f_n}{\omega - E_n + i\delta}. \qquad (3)$$

In Eq. (3) the summation is performed over excited states n between the initial state valence − band and final conduction − band states; $f_n$ is the oscillator strength of the transition:

$$f_n = \frac{2}{3}E_n|\vec{d}_n|^2, \qquad (4)$$

where

$$\vec{d}_n = \langle 0|\vec{r}|n\rangle = \sum_{\vec{k}}\sqrt{\frac{2E_n^0}{E_n}}G_{\vec{k},n}\,\vec{d}_k^{vc} \equiv \sum_{\vec{k}}\sqrt{\frac{2E_n^0}{E_n}}G_{\vec{k},n}\langle\varphi_k^{0v}(\vec{r})|\vec{r}|\varphi_k^{0c}(\vec{r})\rangle \qquad (5)$$

is the transition dipole moment from the ground (0) to the excited state n and $E_n^0$ and $E_n$ are the DFT and TDDFT excitation energies, respectively. The summation in Eq.(5) is performed over all initial and final states, i.e., from the valence to the conduction band for all values of the momentum. Finally, in Eq.(3) we used $\delta = 0.1$. We calculated both TDDFT and DFT absorption spectra (in DFT case one puts in Eq. (5) $E_n \to E_n^0$) and for $G_{\vec{k},n}$ – the eigenvectors obtained by solving Eq.(1) at $F_{\vec{k}\vec{k}'} = 0$).

The mission spectra were obtained from the ab initio absorption spectrum $A(\omega)$ by multiplying it by the Planck factor [35]:

$$E(\omega) = \frac{4\pi\omega^4}{e^{\frac{\omega-\Delta}{T}}-1} A(\omega), \tag{6}$$

where $\Delta$ is the optical gap in the system (the difference between the energy of lowest-excited state and the energy at the top of valence band) and for temperature we used a representative temperature $T = 0.01$ eV (i.e., of order of room temperature). The spectrum at lower temperatures will be similar, but with more narrow peaks.

To obtain an idea about the spatial charge distributions of the electron and hole charges in the exciton, we used the fact that the excited charge density can be expressed in term of the DFT wave functions and Liouville matrix elements as:

$$\delta n(\vec{r},t) = \sum_{k<k_F} \left( |\Psi_{\vec{k}}(\vec{r},t)|^2 - |\Psi_{\vec{k}}(\vec{r},0)|^2 \right)$$

$$= \sum_{k<k_F} \left( \rho_{\vec{k}}^{cv}(t)\varphi_{\vec{k}}^{0c}(\vec{r})\varphi_{\vec{k}}^{0v*}(\vec{r}) + \rho_{\vec{k}}^{vc}(t)\varphi_{\vec{k}}^{0v}(\vec{r})\varphi_{\vec{k}}^{0c*}(\vec{r}) + \rho_{\vec{k}}^{cc}(t)\varphi_{\vec{k}}^{0c}(\vec{r})\varphi_{\vec{k}}^{0c*}(\vec{r}) \right). \tag{7}$$

Thus, the change of the charge density due to creation of the exciton in state n is $\sum_{k<k_F} G_{\vec{k},n}(t)\varphi_{\vec{k}}^{0c}(\vec{r})\varphi_{\vec{k}}^{0v*}(\vec{r})$, where $G_{\vec{k},n}$ are components of the corresponding eigenvector obtained from Eq.(1) for $\rho_{\vec{k}}^{cv}(t)$, and one can write for the excitonic charge density for the state n:

$$|\Psi_{X,n}(r)|^2 = \sum_{k<k_F} G_{\vec{k},n}\varphi_{\vec{k}}^{0c}(\vec{r})\varphi_{\vec{k}}^{0v*}(\vec{r}). \tag{8}$$

For a better visualization of the calculated exciton charge distribution (for the considered state n=1), we plotted individual contributions of the electron and hole charges as $\sum_{k<k_F} G_{\vec{k},1}\varphi_{\vec{k}}^{0c}(r)\varphi_{\vec{k}}^{0v*}(0)$ and $\sum_{k<k_F} G_{k,\vec{1}}\varphi_{\vec{k}}^{0c}(0)\varphi_{\vec{k}}^{0v*}(\vec{r})$ (i.e., by fixing the electron and hole coordinates to be equal zero), correspondingly.

## 3. Results and Discussion

In this section, we summarize the results of our calculations and their analysis. We first focus on the electronic structure of the pristine and hydrogenated single-layer MoS$_2$. This is followed by considerations of the binding energy of the excitons in Sec. 3.2. In Sec. 3.3 we discuss the calculated absorption and emission spectra of the systems of interest which provides a connection with the observed PL data and finally in Sec. 3.4, we examine the details of the charge distribution related to the excitons being investigated. Our conclusions are summarized in Sec. 3.5.

### 3.1 Electronic structure of pristine and hydrogenated single-layer MoS$_2$

We obtained the optimized lattice parameters for the pristine and hydrogenated single layer MoS$_2$, shown schematically in Fig.1, after ionic relaxation using the DFT procedure mentioned above. The optimized lattice constant for single-layer MoS$_2$ is 3.124 and 3.186 Å, with LDA and PBE, respectively, in a good agreement with previous calculations[36]. The band structure obtained using the optimized lattice constant, of single-layer MoS$_2$ is shown in Figure 2. The obtained direct bandgap at the K point is ∼ 1.78 eV and 1.66 eV for LDA-PZ and GGA-PBE, respectively which is also in agreement with previously reported theoretical [37, 38] and experimental values [4, 39] (though, in the experimental case the agreement is accidental, since the measured gap is defined by excitonic states).

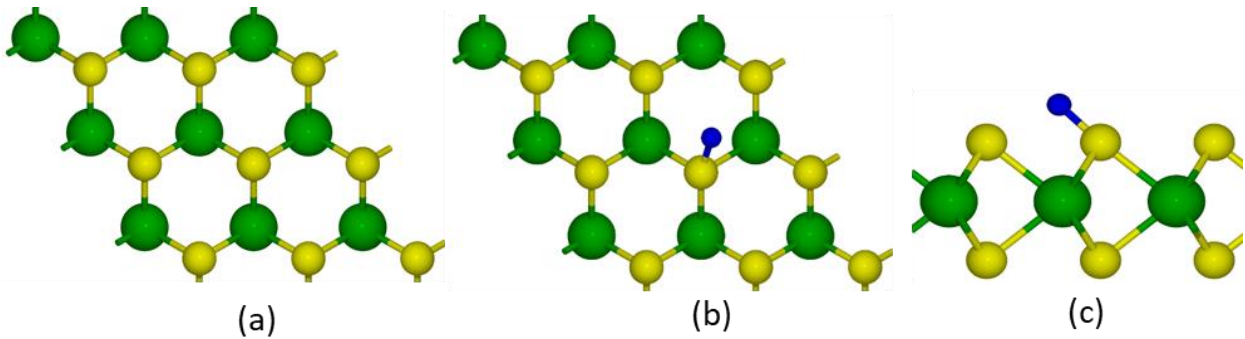

Figure 1: Schematic representation of (a) pristine, top view and (b) top view and (c) side view of hydrogenated MoS$_2$ used in the calculations. Large green balls represent Mo atoms, yellow balls represent sulphur atoms , and blue balls represent hydrogen atoms, respectively.

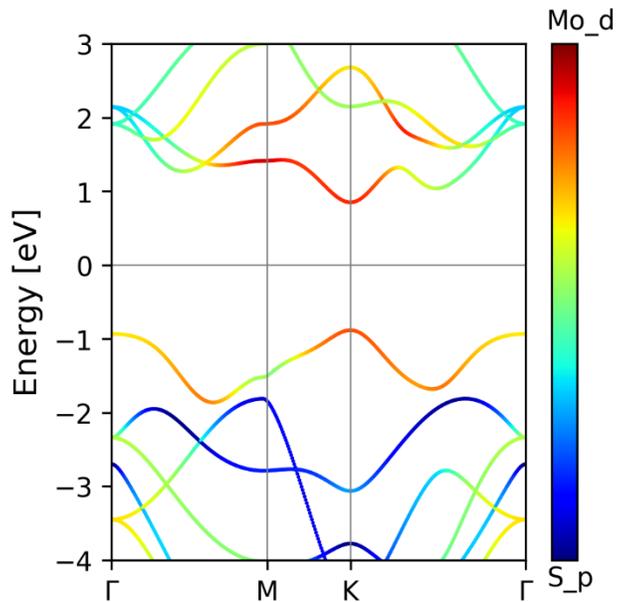

Figure 2: Calculated band structure of pristine single-layer MoS$_2$ obtained by using GGA-PBE. The blue and red colors represent the contribution of the S-*p* and Mo-*d* orbitals, respectively.

Hydrogenation leads to a significant modification of the band structure. We varied the number of H atoms from dilute limit (1/25) to full coverage. In the limiting full coverage case, the system is in the metallic regime (the band structure and the projected DOS are shown in Figures 3a and 4a, respectively), while at low coverages (the hydrogen concentrations 1/9 and 1/25) the system remains gapped and the electronic band structure includes partially-occupied spin-polarized midgap states – occupied spin-up states and empty spin-down states (Figure 5). These spin-polarized states are present in the dilute limit of the hydrogen concentration and are stable at least up to concentration ~10%. In the case of fully hydrogenated MoS$_2$, the system is in a paramagnetic gapped state. We have also performed corresponding calculations with adsorbed Li and Na atoms and found the systems to be in a paramagnetic metal state at all values of the adatom concentration (see Figs. 4b,4c, 5b and 5c, and also Refs.[40, 41]). A possible reason for this is the larger radius of the s-orbitals of Li and Na that is responsible for stronger hybridization of these states with those of MoS$_2$, and hence their delocalization, and diminished effects of spin-polarization.

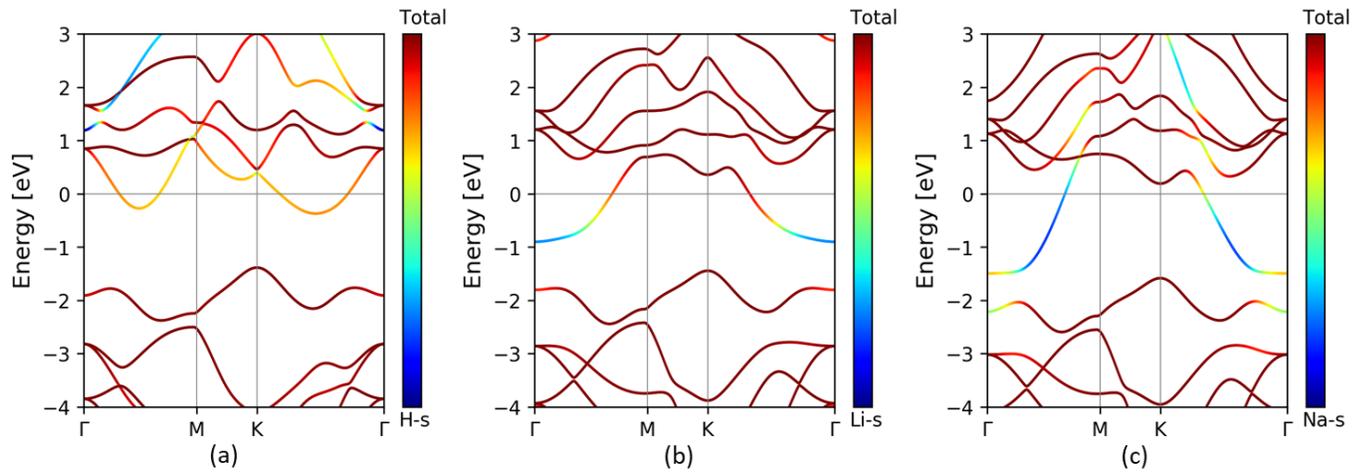

Figure 3. Band structure of MoS$_2$ fully covered with (a) hydrogen (b) lithium (c) sodium, calculated using GGA-PBE. The blue shades represent the contribution of the H, Li, Na-*s* states and red those of Mo and S orbitals. Here and in Figs. 5 and 6, the horizontal black lines mark the Fermi energy.

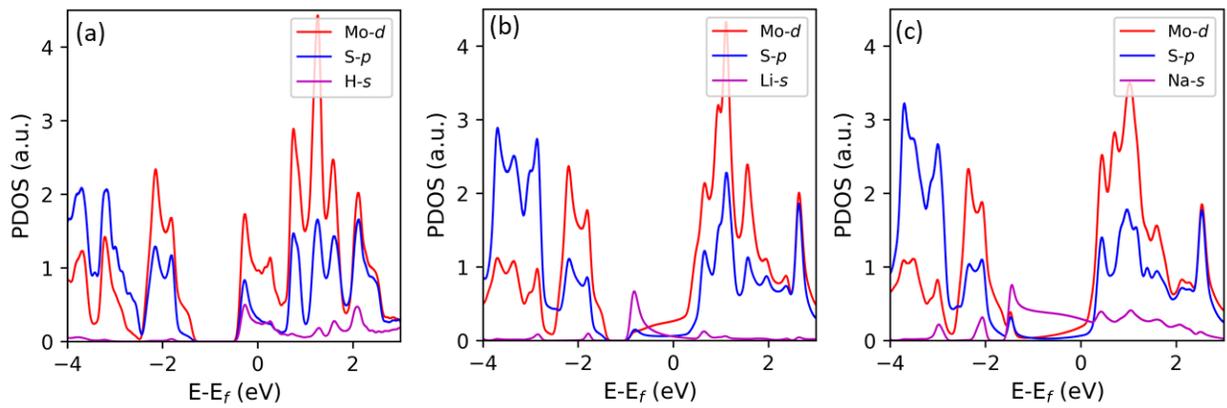

Figure 4: Projected density of states (PDOS) of MoS$_2$ fully-covered with (a) hydrogen, (b) lithium and (c) sodium atoms. The results are obtained with GGA-PBE.

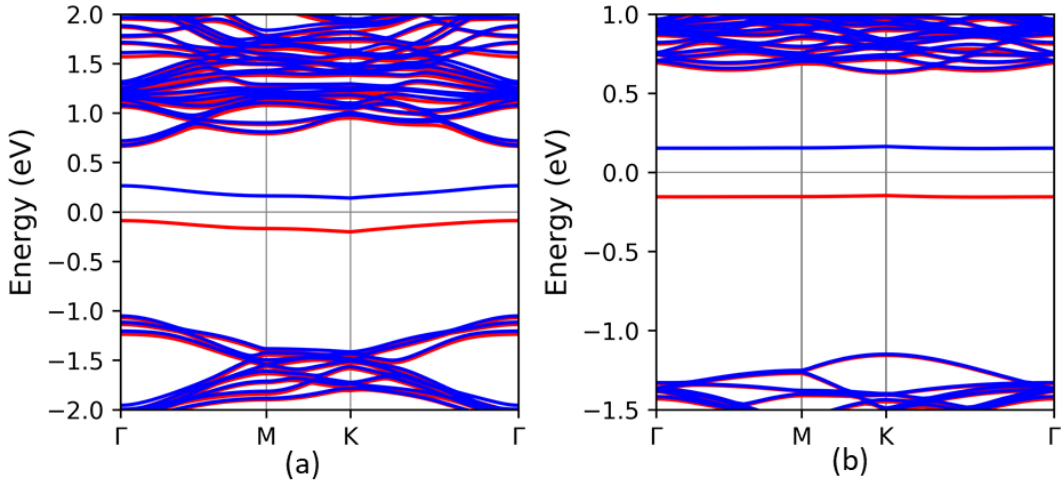

Figure 5. Band structure of hydrogenated MoS$_2$ at (a) 1/9 coverage and (b) 1/25 coverage obtained with GGA-PBE. Spin-up and spin-down states are shown in red and blue colors, correspondingly.

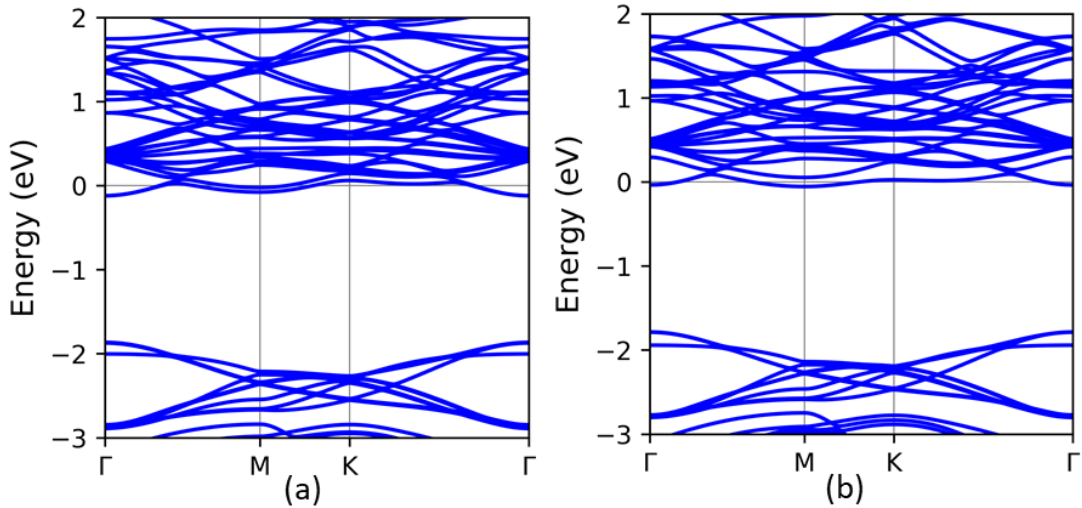

Figure 6. The same as in the previous figure in the case of MoS$_2$ for 1/9 coverage by Li (a) and Na (b) atoms.

## 3.2 Binding energies of excitons

As the next step, we calculated the exciton binding energies for pristine and partially hydrogen covered (1/9 and 1/25) single-layer MoS$_2$. In the pristine system, excitons are formed by conduction-band electrons and valence-band holes (Figure 7). The excitonic binding energy ~1.067eV obtained with the Slater XC kernel is of the same order of magnitude (albeit 2 times larger) when compared with experimental data (0.22-0.57eV)[42, 43]. In the case of hydrogenated systems, the variety of excitonic states is much richer. As it is shown schematically in Figure *8* ,

one may expect bound states of electrons and holes formed by both MoS$_2$ and hydrogen bands. Solving Eq.(1) for such different combinations of electron and hole bands, we have found that many of the corresponding binding energies are rather large, especially for lower (1/25) coverage (Table I).

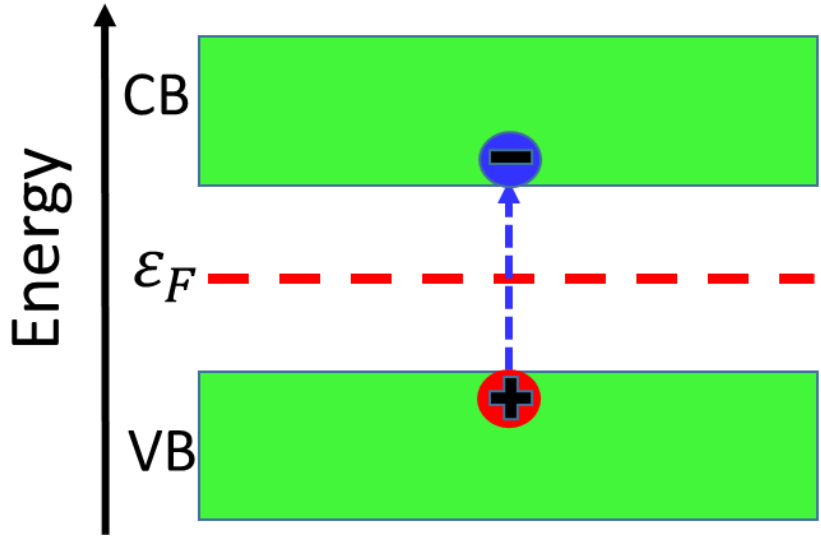

Figure 7:Schematic representation of e-h pair forming the exciton in the non-hydrogenated system.

**Table 1:** The calculated exciton binding energies in meV and the corresponding dipole strengths, using the BEE code, for two coverages of hydrogen. The different types of excitons (A-D) are defined in Fig.8. The dipole strengths are given in units of the strength of the "standard" excitonic state formed by the valence hole and conduction electron (B in Fig.8).

|         | 1/25 Coverage |          |          | 1/9 Coverage |         |          |
|---------|---------|---------|----------|---------|---------|----------|
| Exiton  | Slater  | LR      | Strength | Slater  | LR      | Strength |
| A       | -63.708 | -61.346 | 0.05     | -2.406  | -1.885  | 5        |
| B       | -57.016 | -57.009 | 1        | -1.126  | -0.679  | 1        |
| C       | -82.592 | -97.578 | 47       | -26.771 | -40.25  | 27       |
| D       | -57.043 | -57.053 | 0.1     | -1.054  | -0.834  | 6        |

Smaller binding energies of the "standard" B exciton in the case of higher (1/9) coverage can be explained by enhanced screening effects coming from H electrons, and smaller binding energy of the hydrogen-state exciton, i.e. by hybridization of the "excitonic" hydrogen states with surrounding H atom states that result in a weaker electron-hole interaction. The strongest binding energy and strength was found for the C exciton, when both electron and hole are localized on hydrogen atoms. This can be explained by the local character of the charges that form the exciton. As mentioned above, the strength of this state decreases with increasing coverage. Another important result is the enhanced strength of the mixed excitonic states A and B at 1/9-coverage due to larger hybridization of the hydrogen and $MoS_2$ states at higher coverages (when the hydrogen electrons are more spread out over the surface).

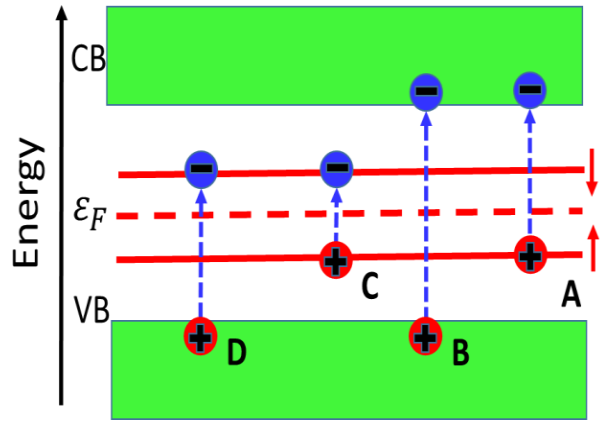

Figure 8: Schematic representation of possible excitonic states in the hydrogenated system.

## 3.3 Absorption and emission spectra

In Figure 9, we plot the DFT and TDDFT (Slater XC kernel) absorption and emission spectra for the pure and the 1/25 hydrogenated $MoS_2$ in the two-band approximation. (In the hydrogenated case, the bands are the occupied and unoccupied mid-gap hydrogen bands).

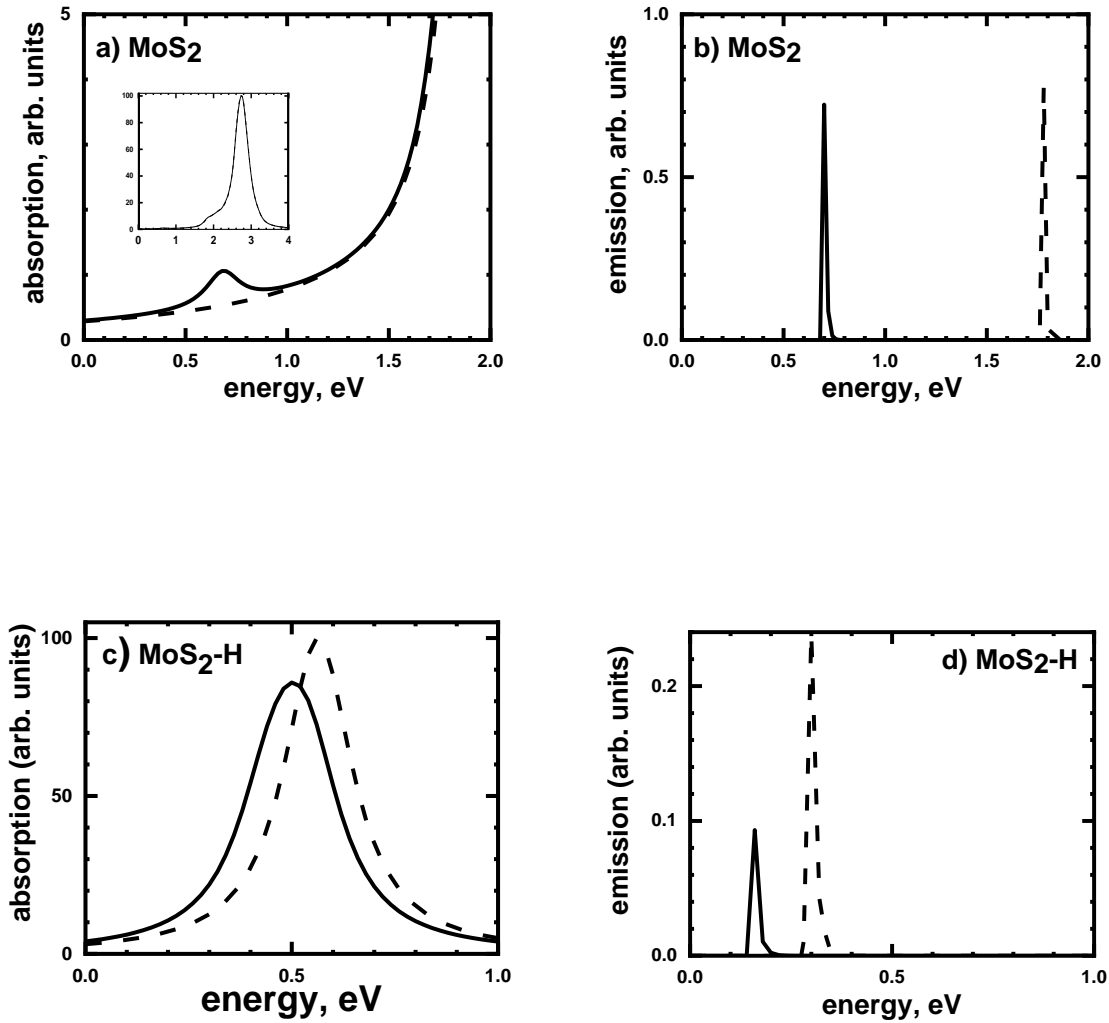

Figure 9: Comparison of TDDFT (solid curves) and DFT (dashed curves) results. Top row: absorption (a) and emission (b) spectra of single-layer $MoS_2$. Bottom row: absorption (c) and emission (d) spectra of the $MoS_2$-H system with 1/25 hydrogen coverage.

As is evident from Figure 9, TDDFT and DFT spectra are rather different, as expected, because of the deficiencies of DFT in proper inclusion of exchange-correlation effects. What is more important, the emission spectrum of the hydrogenated system is in the infra-red frequency range (the peak of the spectrum is at ~0.15eV), while in the pristine system it is in the visible range (by taking into account the fact that the DFT electronic bandgap 1.78eV is ~1eV smaller than the experimental one[42]). The quench of the visible photoluminescence in single-layer $MoS_2$ after hydrogenation is in agreement with experimental observations[17].

## 3.4 Exciton charge distribution

For visualization of the size of the excitonic states, we have calculated charge distributions for the electron and hole states that form valence-conduction and hydrogen-band excitons B and C (the "standard" and the strongest-bound excitons, see Figure *8*) at different coverages. The results are shown in Figure 10. Our calculations find the size of the excitons to be several angstroms for both cases. Importantly, the electron and hole that form hydrogen-band exciton are not localized on the same atom (Fig. 10a) and 10b)).

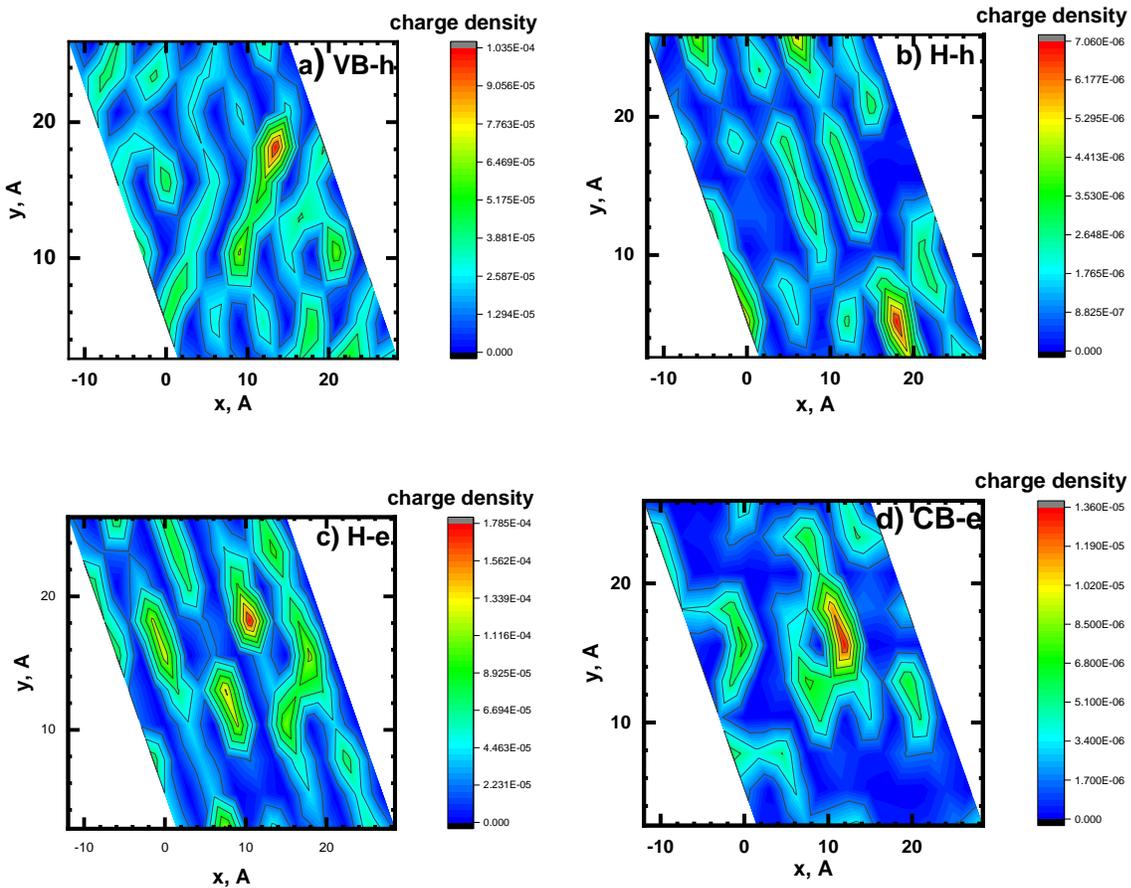

Figure 10: The distribution of the hole (h) and electron (e) charges for the lowest energy excitonic states formed by different band charges: top valence band (VB-h), two hydrogen (initially, one occupied H-h and one empty H-e) bands and bottom conduction band (CB) states for the 1/25 hydrogen density (see the cartoon in Fig. 8 for the states ordering).

## 4. Conclusions

We have studied electronic and excitonic properties of single-layer $MoS_2$ for three hydrogen coverages using a combined DFT and TDDFT approach. As our calculations show, with increasing hydrogen concentration, the system spectrum initially acquires spin-polarized midgap states, and at total coverage the system transforms into a paramagnetic metal. The situation is very different with other s-orbital adsorbates - Li and Na – which prodce a metallic regime at all coveragse. We trace the difference in behavior to a weaker hybridization of H electrons with each other and $MoS_2$ states, owing to the smaller radius of the H-s wave function, which produces localized mid-gap hydrogen states (flat bands).

We have further demonstrated that the spin-polarized states play a very important role in the absorption and excitonic properties of the system, resulting in a dominating hydrogen excitonic state with infrared absorption and emission, in contrast to the visible spectra exhibited by pure $MoS_2$. This result is in agreement with experimental data [17]. Furthermore, our findings provide evidence that hydrogenation is a knob that can be utilized for tuning the absorption spectrum of single-layer $MoS_2$ . The existence of a mid-gap excitonic state in a hydrogenated 2D TMDC which has relatively large binding energy and long lifetime (large dipole strength) is an important finding by itself, one that may have applications in energy harvesting technologies.

Several important questions that remain open. For example, there is a possibility of mid-gap and other higher-order bound states, such as trions and biexcitons, in $MoS_2$-H. These mid-gap states are also expected to have large binding energy owing to a strong charge localization. Another important question concerns the precise character of the ultrafast dynamics of excitons and excited electrons and holes, and finally the time-dependent emissive properties of the system, especially at the femtosecond timescale. We look forward to addressing these issues in the near future, in conjunction with experimental measurements of the same.

Acknowledgements

This work was supported in part by United States Department of Energy under grant No. DE-FG02-07ER46354. We are thankful to Lyman Baker for critical reading of the manuscript and to Duy Le for many useful discussions.

# Bibiliography

1. Furchi, M.M., et al., *Mechanisms of Photoconductivity in Atomically Thin MoS2.* Nano Letters, 2014. **14**(11): p. 6165-6170.
2. Buscema, M., et al., *Photocurrent generation with two-dimensional van der Waals semiconductors.* Chemical Society Reviews, 2015. **44**(11): p. 3691-3718.
3. Lopez-Sanchez, O., et al., *Ultrasensitive photodetectors based on monolayer MoS2.* Nat Nano, 2013. **8**(7): p. 497-501.
4. Splendiani, A., et al., *Emerging Photoluminescence in Monolayer MoS2.* Nano Letters, 2010. **10**(4): p. 1271-1275.
5. Conley, H.J., et al., *Bandgap Engineering of Strained Monolayer and Bilayer MoS2.* Nano Letters, 2013. **13**(8): p. 3626-3630.
6. Li, Y., et al., *Metal to semiconductor transition in metallic transition metal dichalcogenides.* Journal of Applied Physics, 2013. **114**(17): p. 174307.
7. Qu, Y., H. Pan, and C.T. Kwok, *Hydrogenation-controlled phase transition on two-dimensional transition metal dichalcogenides and their unique physical and catalytic properties.* 2016. **6**: p. 34186.
8. Santos, V.P., et al., *Mechanistic Insight into the Synthesis of Higher Alcohols from Syngas: The Role of K Promotion on MoS2 Catalysts.* ACS Catalysis, 2013. **3**(7): p. 1634-1637.
9. Andersen, A., et al., *Adsorption of Potassium on MoS2(100) Surface: A First-Principles Investigation.* The Journal of Physical Chemistry C, 2011. **115**(18): p. 9025-9040.
10. Kotarba, A., et al., *Modification of Electronic Properties of Mo2C Catalyst by Potassium Doping: Impact on the Reactivity in Hydrodenitrogenation Reaction of Indole.* The Journal of Physical Chemistry B, 2004. **108**(9): p. 2885-2892.
11. Cai, Y., et al., *Constructing metallic nanoroads on a MoS 2 monolayer via hydrogenation.* Nanoscale, 2014. **6**(3): p. 1691-1697.
12. Komesu, T., et al., *Occupied and unoccupied electronic structure of Na doped MoS2(0001).* Applied Physics Letters, 2014. **105**(24): p. 241602.
13. Tosun, M., et al., *High-Gain Inverters Based on WSe2 Complementary Field-Effect Transistors.* ACS Nano, 2014. **8**(5): p. 4948-4953.
14. Komesu, T., et al., *Adsorbate doping of MoS2 and WSe2: the influence of Na and Co.* Journal of Physics: Condensed Matter, 2017. **29**(28): p. 285501.
15. Pierucci, D., et al., *Tunable doping in hydrogenated single layered molybdenum disulfide.* ACS nano, 2017. **11**(2): p. 1755-1761.
16. Ma, K.Y., et al., *Hydrogenation of monolayer molybdenum diselenide via hydrogen plasma treatment.* Journal of Materials Chemistry C, 2017. **5**(43): p. 11294-11300.
17. L. Bartels, private communication.
18. Shi, H., et al., *Strong ferromagnetism in hydrogenated monolayer MoS 2 tuned by strain.* Physical Review B, 2013. **88**(20): p. 205305.

19. Jeon, G.W., K.W. Lee, and C.E. Lee, *Ferromagnetism in monolayer MoS2 dictated by hydrogen adsorption sites and concentration.* Physica E: Low-dimensional Systems and Nanostructures, 2018. **104**: p. 309-313.
20. Sorescu, D.C., D.S. Sholl, and A.V. Cugini, *Density functional theory studies of the interaction of H, S, Ni− H, and Ni− S complexes with the MoS2 Basal Plane.* The Journal of Physical Chemistry B, 2004. **108**(1): p. 239-249.
21. Xu, Y., et al., *First-principle study of hydrogenation on monolayer MoS2.* Aip Advances, 2016. **6**(7): p. 075001.
22. Shao, B., et al., *Optically and Electrically Controllable Adatom Spin–orbital Dynamics in Transition Metal Dichalcogenides.* Nano Letters, 2017. **17**(11): p. 6721-6726.
23. Zou, J., et al., *Contrasting properties of hydrogenated and protonated single-layer h-BN from first-principles.* Journal of Physics: Condensed Matter, 2018. **30**(6): p. 065001.
24. Dan, -.W., -.Z. Juan, and -.T. Li-Ming, *- Stability and electronic structure of hydrogenated two-dimensional transition metal dichalcogenides: First-principles study.* - Acta Physica Sinica, 2019. **- 68**(- 3): p. - 037102-1.
25. Savini, G., A. Ferrari, and F. Giustino, *First-principles prediction of doped graphane as a high-temperature electron-phonon superconductor.* Physical review letters, 2010. **105**(3): p. 037002.
26. Loktev, V. and V. Turkowski, *Possible high-temperature superconductivity in multilayer graphane: Can the cuprates be beaten?* Journal of Low Temperature Physics, 2011. **164**(5-6): p. 264-271.
27. Giannozzi, P., et al., *QUANTUM ESPRESSO: a modular and open-source software project for quantum simulations of materials.* Journal of physics: Condensed matter, 2009. **21**(39): p. 395502.
28. Perdew, J.P., K. Burke, and M. Ernzerhof, *Generalized gradient approximation made simple.* Physical review letters, 1996. **77**(18): p. 3865.
29. Perdew, J.P. and A. Zunger, *Self-interaction correction to density-functional approximations for many-electron systems.* Physical Review B, 1981. **23**(10): p. 5048-5079.
30. Turkowski, V., A. Leonardo, and C.A. Ullrich, *Time-dependent density-functional approach for exciton binding energies.* Physical Review B, 2009. **79**(23): p. 233201.
31. Turkowski, V., N.U. Din, and T.S. Rahman, *Time-dependent density-functional theory and excitons in bulk and two-dimensional semiconductors.* Computation, 2017. **5**(3): p. 39.
32. Ramirez-Torres, A., V. Turkowski, and T.S. Rahman, *Time-dependent density-matrix functional theory for trion excitations: Application to monolayer MoS 2 and other transition-metal dichalcogenides.* Physical Review B, 2014. **90**(8): p. 085419.
33. Casida, M.E., *Time-dependent density functional response theory for molecules*, in *Recent Advances In Density Functional Methods: (Part I)*. 1995, World Scientific. p. 155-192.
34. Ruger, R., et al., *Efficient calculation of electronic absorption spectra by means of intensity-selected time-dependent density functional tight binding.* Journal of chemical theory and computation, 2015. **11**(1): p. 157-167.
35. Dresselhaus, M., et al., *Solid State Properties.* 2018.
36. Ramasubramaniam, A., *Large excitonic effects in monolayers of molybdenum and tungsten dichalcogenides.* Physical Review B, 2012. **86**(11): p. 115409.
37. Johari, P. and V.B. Shenoy, *Tuning the Electronic Properties of Semiconducting Transition Metal Dichalcogenides by Applying Mechanical Strains.* ACS Nano, 2012. **6**(6): p. 5449-5456.

38. Ding, Y., et al., *First principles study of structural, vibrational and electronic properties of graphene-like MX2 (M=Mo, Nb, W, Ta; X=S, Se, Te) monolayers.* Physica B: Condensed Matter, 2011. **406**(11): p. 2254-2260.
39. Mak, K.F., et al., *Atomically Thin \${\mathrm{MoS}}_{2}$: A New Direct-Gap Semiconductor.* Physical Review Letters, 2010. **105**(13): p. 136805.
40. Komesu, T., et al., *Occupied and unoccupied electronic structure of Na doped MoS2 (0001).* Applied Physics Letters, 2014. **105**(24): p. 241602.
41. Ersan, F., G.k. Gökoğlu, and E. Aktürk, *Adsorption and diffusion of lithium on monolayer transition metal dichalcogenides (mos2 (1–x) se2 x) alloys.* The Journal of Physical Chemistry C, 2015. **119**(51): p. 28648-28653.
42. Mak, K.F., et al., *Atomically thin MoS 2: a new direct-gap semiconductor.* Physical review letters, 2010. **105**(13): p. 136805.
43. Wu, F., F. Qu, and A.H. Macdonald, *Exciton band structure of monolayer MoS 2.* Physical Review B, 2015. **91**(7): p. 075310.